\documentclass[a4paper,11pt]{article}

\newif\ifjinstmode

\usepackage{jinstpub} 

\usepackage{graphicx} 
\usepackage{subcaption}
\usepackage{hyperref}
\usepackage{orcidlink}
\usepackage[version=4]{mhchem} 

\usepackage{lineno}
\ifjinstmode
    \linenumbers
\fi

\title{Sensitivity enhancement techniques for cryogenic calorimeters in the NUCLEUS experiment}

\abstract{Phonon-mediated cryogenic calorimeters find application in rare event searches due to their intrinsically low energy threshold. Achieving optimal sensitivity for these kinds of detectors is crucial for signal identification, leading to various optimization techniques. In this work, we present two complementary methods to increase the sensitivity of cryogenic detectors read out by transition-edge sensors, developed and tested in the context of the NUCLEUS experiment. The first procedure maps the signal-to-noise ratio of the device across a wide range of operating points, to identify the configuration with maximal sensitivity to be used during data taking. The second method exploits the double readout of the detector, combining the information from different channels with a two-dimensional optimum filter analysis that effectively lowers the energy threshold. By combining both techniques at the same time, we obtain a baseline resolution of $2.94\pm0.05_{\text{stat}}\pm0.74_{\text{syst}}$~eV using a CaWO$_4$ based detector, achieving a promising result ahead of the first run of NUCLEUS at the experimental site.} 

\keywords{Cryogenic detectors, Superconductive detectors, Instrument optimisation, Data processing methods}

\usepackage{xcolor}

\begin{document}




%
\newcount\authorStyle

\authorStyle=0

%
\newcount\instByName

\instByName=0

\setcounter{footnote}{0}
\renewcommand{\thefootnote}{\alph{footnote}}

\newcommand{\iNUCLEUScontactEmail}{Contact E-Mail of NUCLEUS Collaboration : info@nucleus-experiment.org}

\ifcase\authorStyle
    \collaboration{NUCLEUS Collaboration}
\else
\fi

\ifnum\authorStyle=1
\else
    \newcommand{\orgdiv}[1]{#1}%
    \newcommand{\orgname}[1]{#1}%
    \newcommand{\orgaddress}[1]{#1}%
    \newcommand{\street}[1]{#1}%
    \newcommand{\postcode}[1]{#1}%
    \newcommand{\city}[1]{#1}%
    \newcommand{\state}[1]{#1}%
    \newcommand{\country}[1]{#1}%
\fi





\newcommand{\iMBI}{%
    \orgname{Marietta-Blau-Institut f{\"u}r Teilchenphysik der {\"O}sterreichischen Akademie der Wissenschaften}, 
    \orgaddress{
        \street{Dominikanerbastei~16}, 
        \city{Wien}, 
        \postcode{A-1010}, 
        \country{Austria}%
        }%
    }

\newcommand{\iTUW}{%
    \orgdiv{Atominstitut}, 
    \orgname{Technische Universit\"at Wien}, 
    \orgaddress{
        \street{Stadionallee~2}, 
        \city{Wien}, 
        \postcode{A-1020}, 
        \country{Austria}%
        }%
    }


\newcommand{\iCEA}{%
    \orgdiv{IRFU}, 
    \orgname{CEA, Universit\'{e} Paris-Saclay}, 
    \orgaddress{
        \street{B\^{a}timent 141}, 
        \city{Gif-sur-Yvette}, 
        \postcode{F-91191}, 
        \country{France}%
        }%
    }

\newcommand{\iEdF}{%
    \orgdiv{Centre nucl{\'e}aire de production d'{\'}electricit{\'e} de Chooz, Service Automatismes-Essais}, 
    \orgname{{\'E}lectricit{\'e} de France}, 
    \orgaddress{
        \street{}, 
        \city{Givet}, 
        \postcode{F-08600}, 
        \country{France}%
        }%
    }


\newcommand{\iMPIK}{%
    \orgname{Max-Planck-Institut für Kernphysik}, 
    \orgaddress{%
        \street{Saupfercheckweg 1}, 
        \city{Heidelberg}, 
        \postcode{D-69117}, 
        \country{Germany}%
        }%
    }

\newcommand{\iMPP}{%
    \orgname{Max-Planck-Institut f{\"u}r Physik}, 
    \orgaddress{
        \street{Boltzmannstra{\ss}e~8}, 
        \city{Garching}, 
        \postcode{D-85748}, 
        \country{Germany}%
        }%
    }

\newcommand{\iTUM}{%
    \orgdiv{Physik-Department, TUM School of Natural Sciences}, 
    \orgname{Technische Universit\"at M\"unchen}, 
    \orgaddress{
        \street{James-Franck-Straße 1}, 
        \city{Garching}, 
        \postcode{D-85748}, 
        \country{Germany}%
        }%
    }


\newcommand{\iINFNRoma}{%
    \orgname{Istituto Nazionale di Fisica Nucleare -- Sezione di Roma}, 
    \orgaddress{
        \street{Piazzale Aldo Moro 2}, 
        \city{Roma}, 
        \postcode{I-00185}, 
        \country{Italy}%
        }%
    }

\newcommand{\iSapienza}{%
    \orgdiv{Dipartimento di Fisica}, 
    \orgname{Sapienza Universit\`{a} di Roma}, 
    \orgaddress{
        \street{Piazzale Aldo Moro 5}, 
        \city{Roma}, 
        \postcode{I-00185}, 
        \country{Italy}%
        }%
    }
    
\newcommand{\iINFNTorVergata}{%
    \orgname{Istituto Nazionale di Fisica Nucleare -- Sezione di Roma "Tor Vergata"}, 
    \orgaddress{
        \street{Via della Ricerca Scientifica 1}, 
        \city{Roma}, 
        \postcode{I-00133}, 
        \country{Italy}%
        }%
    }

\newcommand{\iTorVergata}{%
    \orgdiv{Dipartimento di Fisica}, 
    \orgname{Universit\`{a} di Roma "Tor Vergata"}, 
    \orgaddress{
        \street{Via della Ricerca Scientifica 1}, 
        \city{Roma}, 
        \postcode{I-00133}, 
        \country{Italy}%
        }%
    }

\newcommand{\iCNR}{%
    \orgdiv{Istituto di Nanotecnologia}, 
    \orgname{Consiglio Nazionale delle Ricerche}, 
    \orgaddress{
        \street{Piazzale Aldo Moro 5}, 
        \city{Roma}, 
        \postcode{I-00185}, 
        \country{Italy}%
        }%
    }

\newcommand{\iINFNFerrara}{%
    \orgname{Istituto Nazionale di Fisica Nucleare -- Sezione di Ferrara}, 
    \orgaddress{
        \street{Via Giuseppe Saragat 1c}, 
        \city{Ferrara}, 
        \postcode{I-44122}, 
        \country{Italy}%
        }%
    }

\newcommand{\iFerrara}{%
    \orgdiv{Dipartimento di Fisica}, 
    \orgname{Universit{\`a} di Ferrara}, 
    \orgaddress{
        \street{Via Giuseppe Saragat 1}, 
        \city{Ferrara}, 
        \postcode{I-44122}, 
        \country{Italy}%
        }%
    }

\newcommand{\iINFNLnGS}{%
    \orgname{Istituto Nazionale di Fisica Nucleare -- Laboratori Nazionali del Gran Sasso}, 
    \orgaddress{
        \street{Via Giovanni Acitelli 22}, 
        \city{Assergi (L’Aquila)}, 
        \postcode{I-67100}, 
        \country{Italy}%
        }%
    }

\newcommand{\iBicocca}{%
    \orgdiv{Dipartimento di Fisica}, 
    \orgname{Universit\`{a} di Milano Bicocca}, 
    \orgaddress{
        \street{}, 
        \city{Milan}, 
        \postcode{I-20126}, 
        \country{Italy}%
        }%
    }


\newcommand{\iCoimbra}{%
    \orgdiv{LIBPhys-UC, Departamento de Fisica}, 
    \orgname{Universidade de Coimbra}, 
    \orgaddress{
        \street{Rua Larga 3004-516}, 
        \city{Coimbra}, 
        \postcode{P3004-516}, 
        \country{Portugal}%
        }%
    }

%
%
\newcommand{\iAlsoAtCoimbra}{Also at \iCoimbra}
\newcommand{\iNowAtMPIK}{Now at \iMPIK}
\newcommand{\iNowAtLNGS}{Now at \iINFNLnGS}
\newcommand{\iNowAtKiutra}{Now at kiutra GmbH, Fl{\"o}{\ss}ergasse 2, D-81369 Munich, Germany}


\ifnum\instByName=1
    
    \newcommand{\MBI}{MBI}
    \newcommand{\TUW}{TUW}

    \newcommand{\CEA}{CEA}
    \newcommand{\EdF}{EdF}

    \newcommand{\MPIK}{MPIK}
    \newcommand{\MPP}{MPP}
    \newcommand{\TUM}{TUM}

    \newcommand{\INFNRoma}{INFNRoma}
    \newcommand{\Sapienza}{Sapienza}
    \newcommand{\INFNTorVergata}{INFNTorVergata}
    \newcommand{\TorVergata}{TorVergata}
    \newcommand{\CNR}{CNR}
    \newcommand{\INFNFerrara}{INFNFerrara}
    \newcommand{\Ferrara}{Ferrara}
    \newcommand{\INFNLnGS}{INFNLnGS}
    \newcommand{\Bicocca}{Bicocca}    

    \newcommand{\Coimbra}{Coimbra}
\else    
    
    \newcommand{\MBI}{3}
    \newcommand{\TUW}{1}

    \newcommand{\CEA}{5}
    \newcommand{\EdF}{ERROR}

    \newcommand{\MPIK}{ERROR}
    \newcommand{\MPP}{2}
    \newcommand{\TUM}{8}

    \newcommand{\INFNRoma}{6}
    \newcommand{\Sapienza}{7}
    
    \newcommand{\INFNTorVergata}{4}
    \newcommand{\TorVergata}{9}
    
    \newcommand{\CNR}{ERROR}
    
    \newcommand{\Ferrara}{10}
    \newcommand{\INFNFerrara}{11}
    
    \newcommand{\INFNLnGS}{ERROR}
    
    \newcommand{\Bicocca}{ERROR}    

    \newcommand{\Coimbra}{ERROR}

\fi




\ifcase\authorStyle
    \author[\INFNRoma, \Sapienza]{M.~Cappelli~\orcidlink{0009-0002-6148-5964}~\footnote{Corresponding author: M.~Cappelli,~\texttt{matteo.cappelli@roma1.infn.it}}~}
\or
    \author[\Sapienza, \INFNRoma]{
        \fnm{M.}
        \sur{Cappelli}
        \emailAdd{\newline Cappelli, M.: matteo.cappelli@roma1.infn.it}
        \orcidlink{0009-0002-6148-5964}
        }
\else
\fi

\ifcase\authorStyle
    \author[\TUM]{A.~Wallach~\orcidlink{ 0009-0009-1703-9634}~\footnote{Corresponding author: A.~Wallach,~\texttt{alexander.wallach@tum.de}}~}
\or
    \author[\TUM]{
        \fnm{A.} 
        \sur{Wallach} 
        \emailAdd{\newline Wallach, A.: alexander.wallach@tum.de}  
        \orcidlink{0009-0009-1703-9634}
        }
\else
\fi

\ifcase\authorStyle
    \author[\TUW]{\newline H.~Abele~\orcidlink{0000-0002-6832-9051}}
    \affiliation[\TUW]{\iTUW}
\or
    \author[\TUW]{
        \fnm{H.} 
        \sur{Abele} 
        \emailAdd{\newline Abele, H.: hartmut.abele@tuwien.ac.at}  
        \orcidlink{0000-0002-6832-9051}
        }
\else
\fi

\ifcase\authorStyle
    \author[\MPP]{G.~Angloher}
    \affiliation[\MPP]{\iMPP}
\or
    \author[\MPP]{
    \fnm{G.} 
    \sur{Angloher} 
     \emailAdd{\newline Angloher, G.: gangloher@epo.org}
    }
\else
\fi

\ifcase\authorStyle
    \author[\MBI]{B.~Arnold}
    \affiliation[\MBI]{\iMBI}
\or
    \author[\MBI]{
        \fnm{B.}
        \sur{Arnold}
        \emailAdd{\newline Arnold B.: bernhard.arnold@oeaw.ac.at}
        }
\else
\fi
        
\ifcase\authorStyle
    \author[\INFNTorVergata]{M.~Atzori~Corona~\orcidlink{0000-0001-5092-3602}}
    \affiliation[\INFNTorVergata]{\iINFNTorVergata}
\or
    \author[\INFNTorVergata]{
    \fnm{M.}
    \sur{Atzori~Corona}
    \emailAdd{\newline Atzori~Corona, M.: mcorona@roma2.infn.it}
    \orcidlink{0000-0001-5092-3602}
    }
\else
\fi

\ifcase\authorStyle
    \author[\MPP]{A.~Bento~\orcidlink{0000-0002-3817-6015}~\footnote{\iAlsoAtCoimbra}~}
\or
    \author[\MPP]{
        \fnm{A.}
        \sur{Bento}
        \emailAdd{\newline Bento A.: bento@mpp.mpg.de}
        \orcidlink{0000-0002-3817-6015}
        \textsuperscript{\symAlsoAtCoimbra,\,}%
        }
    \else
\fi

\ifcase\authorStyle
    \author[\CEA]{E.~Bossio~\orcidlink{0000-0001-9304-1829}}
    \affiliation[\CEA]{\iCEA}
\or
    \author[\CEA]{
        \fnm{E.}
        \sur{Bossio}
        \emailAdd{\newline Bossio E.: elisabetta.bossio@cea.fr}
        \orcidlink{0000-0001-9304-1829}
        }
    \else
\fi
    
\ifcase\authorStyle
    \author[\MBI]{F.~Buchsteiner}
\or
    \author[\MBI]{
        \fnm{F.}
        \sur{Buchsteiner}
        \emailAdd{\newline Buchsteiner, F. florian.buchsteiner@oeaw.ac.at}
        }
    \else
\fi
    
\ifcase\authorStyle
    \author[\MBI]{J.~Burkhart~\orcidlink{0000-0002-1989-7845}}
\or
    \author[\MBI]{
        \fnm{J.}
        \sur{Burkhart}
        \emailAdd{\newline Burkhart, J.: jens.burkhart@cern.ch}
        \orcidlink{0000-0002-1989-7845}
        }
\else
\fi
    

\ifcase\authorStyle
    \author[\INFNRoma]{F.~Cappella~\orcidlink{0000-0003-0900-6794}}
    \affiliation[\INFNRoma]{\iINFNRoma}
\or
    \author[\INFNRoma]{
        \fnm{F.}
        \sur{Cappella}
        \emailAdd{\newline Cappella, F.: fabio.cappella@roma1.infn.it}
        \orcidlink{0000-0003-0900-6794}
        }
\else
\fi
        

\ifcase\authorStyle
    \author[\INFNRoma]{N.~Casali~\orcidlink{0000-0003-3669-8247}}
\or
    \author[\INFNRoma]{
        \fnm{N.}
        \sur{Casali}
        \emailAdd{\newline Casali, N.: nicola.casali@roma1.infn.it}
        \orcidlink{0000-0003-3669-8247}
        }
\else
\fi
    
\ifcase\authorStyle
    \author[\INFNTorVergata]{R.~Cerulli~\orcidlink{0000-0003-2051-3471}}
\or
    \author[\INFNTorVergata]{
        \fnm{R.}
        \sur{Cerulli}
        \emailAdd{\newline Cerulli, R.: riccardo.cerulli@roma2.infn.it}
        \orcidlink{0000-0003-2051-3471}
        }
\else
\fi
    

\ifcase\authorStyle
    \author[\INFNRoma]{A.~Cruciani~\orcidlink{0000-0003-2247-8067}}
\or
    \author[\INFNRoma]{
        \fnm{A.}
        \sur{Cruciani}
        \emailAdd{\newline Cruciani, A.: angelo.cruciani@roma1.infn.it}
        \orcidlink{0000-0003-2247-8067}
        }
\else
\fi
    
\ifcase\authorStyle
    \author[\INFNRoma]{G.~Del~Castello~\orcidlink{0000-0001-7182-358X}}
\or
    \author[\INFNRoma]{
        \fnm{G.}
        \sur{Del~Castello}
        \emailAdd{\newline Del~Castello, G.: giorgio.delcastello@roma1.infn.it}
        \orcidlink{0000-0001-7182-358X}
        }
\else
\fi

\ifcase\authorStyle
    \author[\INFNRoma, \Sapienza]{M.~del~Gallo~Roccagiovine}
    \affiliation[\Sapienza]{\iSapienza}
\or
    \author[\Sapienza, \INFNRoma]{
        \fnm{M.}
        \sur{del~Gallo~Roccagiovine}
        \emailAdd{\newline del~Gallo~Roccagiovine, M.: matteo.delgalloroccagiovine@roma1.infn.it}
        }
\else
\fi
    

\ifcase\authorStyle
    \author[\TUW]{S.~Dorer~\orcidlink{0009-0001-1670-5780}}
\or
    \author[\TUW]{
        \fnm{S.}
        \sur{Dorer}
        \emailAdd{\newline Dorer, S.: sebastian.dorer@tuwien.ac.at}
        \orcidlink{0009-0001-1670-5780}
        }
\else
\fi
    
\ifcase\authorStyle
    \author[\TUM]{A.~Erhart~\orcidlink{0000-0002-8721-177X}}
    \affiliation[\TUM]{\iTUM}
\or
    \author[\TUM]{
        \fnm{A.}
        \sur{Erhart}
        \emailAdd{\newline Erhart, A.: andreas.erhart@tum.de}
        \orcidlink{0000-0002-8721-177X}
        }
\else
\fi
    
\ifcase\authorStyle
    \author[\MBI]{M.~Friedl~\orcidlink{0000-0002-7420-2559}}
\or
    \author[\MBI]{
        \fnm{M.}
        \sur{Friedl}
        \emailAdd{\newline Friedl, M.: markus.friedl@oeaw.ac.at}
        \orcidlink{0000-0002-7420-2559}}
\else
\fi

\ifcase\authorStyle
    \author[\MBI]{S.~Fichtinger}
\or
    \author[\MBI]{
        \fnm{S.}
        \sur{Fichtinger}
        \emailAdd{\newline Fichtinger, S.: stephan.fichtinger@oeaw.ac.at}
        }
\else
\fi


\ifcase\authorStyle
    \author[\MBI]{V.M.~Ghete~\orcidlink{0000-0002-9595-6560}}
\or
    \author[\MBI]{
        \fnm{V.M.}
        \sur{Ghete}
        \emailAdd{\newline Ghete, V. M.: Vasile-Mihai.Ghete@oeaw.ac.at}
        \orcidlink{0000-0002-9595-6560}
        }
\else
\fi

\ifcase\authorStyle
    \author[\INFNTorVergata, \TorVergata]{M.~Giammei~\orcidlink{0009-0006-9104-2055}}
    \affiliation[\TorVergata]{\iTorVergata}
\or
    \author[\TorVergata, \INFNTorVergata]{
        \fnm{M.}
        \sur{Giammei}
        \emailAdd{\newline Giammei, M: giammeim@roma2.infn.it}
        \orcidlink{0009-0006-9104-2055}
        }
\else
\fi

\ifcase\authorStyle
    \author[\CEA]{C.~Goupy~\orcidlink{0000-0003-4954-5311}~\footnote{\iNowAtMPIK}~}
\or
    \author[\CEA]{
        \fnm{C.} 
        \sur{Goupy} 
        \emailAdd{\newline Goupy, C.: Chloe.Goupy@mpi-hd.mpg.de}  
        \orcidlink{0000-0003-4954-5311}
        \textsuperscript{\symNowAtMPIK,\,}%
        }
\else
\fi


\ifcase\authorStyle
    \author[\MBI]{J.~Hakenm{\"u}ller~\orcidlink{0000-0003-0470-3320}}
\or
    \author[\MBI]{
        \fnm{J.}
        \sur{Hakenm{\"u}ller}
        \emailAdd{\newline Hakenm{\"u}ller, J.: JaninaDorin.Hakenmueller@oeaw.ac.at}
        \orcidlink{0000-0003-0470-3320}}
\else
\fi

\ifcase\authorStyle
    \author[\MPP, \TUM]{D.~Hauff}
\or
    \author[\MPP, \TUM]{
        \fnm{D.}
        \sur{Hauff}
        \emailAdd{\newline Hauff, D.: hauff@mpp.mpg.de}
        }
\else
\fi
    
\ifcase\authorStyle
    \author[\CEA]{F.~Jeanneau~\orcidlink{0000-0002-6360-6136}}
\or
    \author[\CEA]{
        \fnm{F.}
        \sur{Jeanneau}
        \emailAdd{\newline Jeanneau, F.: fabien.jeanneau@cea.fr}
        \orcidlink{0000-0002-6360-6136}}
\else
\fi

\ifcase\authorStyle
    \author[\TUW]{E.~Jericha~\orcidlink{0000-0002-8663-0526}}
\or
    \author[\TUW]{
        \fnm{E.}
        \sur{Jericha}
        \emailAdd{\newline Jericha, E.: erwin.jericha@tuwien.ac.at}~\orcidlink{0000-0002-8663-0526}
        }
\else
\fi

\ifcase\authorStyle
    \author[\TUM]{M.~Kaznacheeva~\orcidlink{0000-0002-2712-1326}}
\or
    \author[\TUM]{
        \fnm{M.}
        \sur{Kaznacheeva}
        \emailAdd{\newline Kaznacheeva, M.: margarita.kaznacheeva@tum.de}
        \orcidlink{0000-0002-2712-1326}
        }
\else
\fi


\ifcase\authorStyle
    \author[\MBI]{H.~Kluck~\orcidlink{0000-0003-3061-3732}}
\or
\author[\MBI]{
    \fnm{H.} 
    \sur{Kluck} 
    \emailAdd{\newline Kluck, H.: Holger.Kluck@oeaw.ac.at}  
    \orcidlink{0000-0003-3061-3732}
    }
\else
\fi

\ifcase\authorStyle
    \author[\MPP]{A.~Langenk{\"a}mper}
\or
    \author[\MPP]{
        \fnm{A.} 
        \sur{Langenk\"{a}mper} 
        \emailAdd{\newline Langenk\"{a}mper, A.: langenk@mpp.mpg.de}  
        \orcidlink{}
        }
\else
\fi

\ifcase\authorStyle
    \author[\CEA, \TUM]{T.~Lasserre~\orcidlink{0000-0002-4975-2321}~\footnote{\iNowAtMPIK}~}
\or
    \author[\CEA, \TUM]{
        \fnm{T.} 
        \sur{Lasserre} 
        \emailAdd{\newline Lasserre, T.: thierry.lasserre@mpi-hd.mpg.de}  
        \orcidlink{0000-0002-4975-2321}
        \textsuperscript{\symNowAtMPIK,\,}%
        }
\else
\fi

\ifcase\authorStyle
    \author[\CEA]{D.~Lhuillier~\orcidlink{0000-0003-2324-0149}}
\or
    \author[\CEA]{
        \fnm{D.} 
        \sur{Lhuillier} 
        \emailAdd{\newline Lhuillier, D.: david.lhuillier@cea.fr}  
        \orcidlink{0000-0003-2324-0149}
        }
\else
\fi

\ifcase\authorStyle
    \author[\MPP]{M.~Mancuso~\orcidlink{0000-0001-9805-475X}}
\or
    \author[\MPP]{
        \fnm{M.} 
        \sur{Mancuso} 
        \emailAdd{\newline Mancuso, M.: michele.mancuso@mpp.mpg.de}  
        \orcidlink{0000-0001-9805-475X}
        }
\else
\fi

\ifcase\authorStyle
    \author[\TUW, \CEA]{R.~Martin}
\or
    \author[\CEA, \TUW]{
        \fnm{R.} 
        \sur{Martin} 
        \emailAdd{\newline Martin,R.: romain.martin2@cea.fr}  
        }
\else
\fi

\ifcase\authorStyle
    \author[\MPP]{B.~Mauri}
\or
    \author[\MPP]{
        \fnm{B.} 
        \sur{Mauri} 
        \emailAdd{\newline Mauri, B.: bmauri@mpp.mpg.de}  
        }
\else
\fi

\ifcase\authorStyle
    \author[\Ferrara, \INFNFerrara]{A.~Mazzolari}
    \affiliation[\Ferrara]{\iFerrara}
    \affiliation[\INFNFerrara]{\iINFNFerrara}
\or
    \author[\Ferrara, \INFNFerrara]{
        \fnm{A.} 
        \sur{Mazzolari} 
        \emailAdd{\newline Mazzolari, A.: andrea.mazzolari@unife.it}  
        }
\else
\fi

\ifcase\authorStyle
    \author[\CEA]{L.~McCallin}
\or
    \author[\CEA]{
        \fnm{L.} 
        \sur{McCallin} 
        \emailAdd{\newline McCallin, L.: liliane.mccallin@cea.fr}  
        }
\else
\fi


\ifcase\authorStyle
    \author[\CEA]{H.~Neyrial}
\or
    \author[\CEA]{
        \fnm{H.} 
        \sur{Neyrial} 
        \emailAdd{\newline Neyrial, H.: Hubert.NEYRIAL@cea.fr}  
        }
\else
\fi

\ifcase\authorStyle
    \author[\CEA]{C.~Nones}
\or
    \author[\CEA]{
        \fnm{C.} 
        \sur{Nones} 
        \emailAdd{\newline Nones, C.: Claudia.Nones@cea.fr}  
        }
\else
\fi

\ifcase\authorStyle
    \author[\TUM]{L.~Oberauer}
\or
    \author[\TUM]{
        \fnm{L.} 
        \sur{Oberauer} 
        \emailAdd{\newline Oberauer, L.: lothar.oberauer@tum.de}  
        }
\else
\fi



\ifcase\authorStyle
    \author[\CEA, \TUM]{L.~Peters~\orcidlink{0000-0002-1649-8582}~\footnote{\iNowAtMPIK}~}
\or
    \author[\TUM, \CEA]{
        \fnm{L.} 
        \sur{Peters} 
        \emailAdd{\newline Peters, L.: lilly.peters@tum.de}  
        \orcidlink{0000-0002-1649-8582}
        \textsuperscript{\symNowAtMPIK,\,}%
        }
\else
\fi

\ifcase\authorStyle
    \author[\MPP]{F.~Petricca~\orcidlink{0000-0002-6355-2545}}
\or
    \author[\MPP]{
        \fnm{F.} 
        \sur{Petricca} 
        \emailAdd{\newline Petricca, F.: petricca@mpp.mpg.de}  
        \orcidlink{0000-0002-6355-2545}
        }
\else
\fi

\ifcase\authorStyle
    \author[\TUM]{W.~Potzel}
\or
    \author[\TUM]{
        \fnm{W.} 
        \sur{Potzel} 
        \emailAdd{\newline Potzel, W.: walter.potzel@tum.de}  
        }
\else
\fi

\ifcase\authorStyle
    \author[\MPP]{F.~Pr\"{o}bst}
\or
    \author[\MPP]{
        \fnm{F.} 
        \sur{Pr\"{o}bst} 
        \emailAdd{\newline Pr\"{o}bst, F.: proebst@mpp.mpg.de}  
        }
\else
\fi

\ifcase\authorStyle
    \author[\MPP]{F.~Pucci~\orcidlink{0000-0003-3782-2393}~\footnote{Now at \iINFNLnGS}~}
\or
    \author[\MPP]{
        \fnm{F.} 
        \sur{Pucci} 
        \emailAdd{\newline Pucci, F.: francesca.pucci@lngs.infn.it} 
        \orcidlink{0000-0003-3782-2393}
        \textsuperscript{\symNowAtLNGS,\,}%
        }
\else
\fi

\ifcase\authorStyle
    \author[\TUW, \MBI]{F.~Reindl~\orcidlink{0000-0003-0151-2174}}
\or
    \author[\TUW, \MBI]{
        \fnm{F.} 
        \sur{Reindl} 
        \emailAdd{\newline Reindl, F.: florian.reindl@tuwien.ac.at}  
        \orcidlink{0000-0003-0151-2174}
        }
\else
\fi


\ifcase\authorStyle
    \author[\Ferrara, \INFNFerrara]{M.~Romagnoni}
\or
    \author[\Ferrara, \INFNFerrara]{
        \fnm{M.} 
        \sur{Romagnoni} 
        \emailAdd{\newline Romagnoni, M.: romagnoni@fe.infn.it}  
        }
\else
\fi

\ifcase\authorStyle
    \author[\TUM]{J.~Rothe~\orcidlink{0000-0001-5748-7428}~\footnote{Now at \iNowAtKiutra}~}
\or
    \author[\TUM]{
        \fnm{J.} 
        \sur{Rothe} 
        \emailAdd{\newline Rothe, J.: johannes.rothe@tum.de}  
        \orcidlink{0000-0001-5748-7428}
        \textsuperscript{\symNowAtKiutra,\,}%
        }
\else
\fi

\ifcase\authorStyle
    \author[\TUM]{N.~Schermer~\orcidlink{0009-0004-4213-5154}}
\or
    \author[\TUM]{
        \fnm{N.} 
        \sur{Schermer} 
        \emailAdd{\newline Schermer, N.: nicole.schermer@tum.de}  
        \orcidlink{0009-0004-4213-5154}
        }
\else
\fi

\ifcase\authorStyle
    \author[\TUW, \MBI]{J.~Schieck~\orcidlink{0000-0002-1058-8093}}
\or
    \author[\MBI, \TUW]{
        \fnm{J.} 
        \sur{Schieck} 
        \emailAdd{\newline Schieck, J.: Jochen.Schieck@oeaw.ac.at}  
        \orcidlink{0000-0002-1058-8093}
        }
\else
\fi

\ifcase\authorStyle
    \author[\TUM]{S.~Sch\"{o}nert~\orcidlink{0000-0001-5276-2881}}
\or
    \author[\TUM]{
        \fnm{S.} 
        \sur{Sch\"{o}nert} 
        \emailAdd{\newline Sch\"{o}nert, S.: schoenert@ph.tum.de}  
        \orcidlink{0000-0001-5276-2881}
        }
\else
\fi

\ifcase\authorStyle
    \author[\TUW, \MBI]{C.~Schwertner}
\or
    \author[\MBI, \TUW]{
        \fnm{C.} 
        \sur{Schwertner} 
        \emailAdd{\newline Schwertner, C.: Christoph.Schwertner@oeaw.ac.at}  
        }
\else
\fi

\ifcase\authorStyle
    \author[\CEA]{L.~Scola}
\or
    \author[\CEA]{
        \fnm{L.} 
        \sur{Scola} 
        \emailAdd{\newline Scola, L.: loris.scola@cea.fr}  
        }
\else
\fi

\ifcase\authorStyle
    \author[\CEA]{G.~Soum-Sidikov~\orcidlink{0000-0003-1900-1794}}
\or
    \author[\CEA]{
        \fnm{G.} 
        \sur{Soum-Sidikov} 
        \emailAdd{\newline Soum-Sidikov, G.: gabrielle.soum@cea.fr}  
        \orcidlink{0000-0003-1900-1794}
        }
\else
\fi

\ifcase\authorStyle
    \author[\MPP]{L.~Stodolsky}
\or
    \author[\MPP]{
        \fnm{L.} 
        \sur{Stodolsky} 
        \emailAdd{\newline Stodolsky, L.: les@mpp.mpg.de}  
        }
\else
\fi

\ifcase\authorStyle
    \author[\TUM]{A.~Schr{\"o}der~\orcidlink{0009-0005-1598-1635}}
\or
    \author[\TUM]{
        \fnm{A.} 
        \sur{Schr{\"o}der} 
        \emailAdd{\newline Schr{\"o}der, A.: alexandra.schroeder@tum.de}  
        \orcidlink{0009-0005-1598-1635}
        }
\else
\fi

\ifcase\authorStyle
    \author[\TUM]{R.~Strauss~\orcidlink{0000-0002-5589-9952}}
\or
    \author[\TUM]{
        \fnm{R.} 
        \sur{Strauss} 
        \emailAdd{\newline Strauss, R.: raimund.strauss@tum.de}  
        \orcidlink{0000-0002-5589-9952}
        }
\else
\fi


\ifcase\authorStyle
    \author[\MBI]{R.~Thalmeier~\orcidlink{0009-0003-4480-0990}}
\or
    \author[\MBI]{
        \fnm{R.} 
        \sur{Thalmeier} 
        \emailAdd{\newline Thalmeier, R.: richard.thalmeier@oeaw.ac.at}  
        \orcidlink{0009-0003-4480-0990}
        }
\else
\fi

\ifcase\authorStyle
    \author[\INFNRoma]{C.~Tomei}
\or
    \author[\INFNRoma]{
        \fnm{C.} 
        \sur{Tomei} 
        \emailAdd{\newline Tomei, C.: claudia.tomei@roma1.infn.it}  
        }
\else
\fi

\ifcase\authorStyle
    \author[\MBI]{L.~Valla~\orcidlink{0009-0003-7140-9196}}
\or
    \author[\MBI]{
        \fnm{L.} 
        \sur{Valla} 
        \emailAdd{\newline Valla, L.: lorenzo.valla@oeaw.ac.at}  
        \orcidlink{0009-0003-7140-9196}
        }
\else
\fi

\ifcase\authorStyle
    \author[\INFNRoma, \Sapienza]{M.~Vignati~\orcidlink{0000-0002-8945-1128}}
\or
    \author[\Sapienza, \INFNRoma]{
        \fnm{M.} 
        \sur{Vignati} 
        \emailAdd{\newline Vignati, M.: Marco.Vignati@roma1.infn.it}  
        \orcidlink{0000-0002-8945-1128}
        }
\else
\fi

\ifcase\authorStyle
    \author[\CEA]{M.~Vivier~\orcidlink{0000-0003-2199-0958}}
\or
    \author[\CEA]{
        \fnm{M.} 
        \sur{Vivier} 
        \emailAdd{\newline Vivier, M.: Matthieu.Vivier@cea.fr}  
        \orcidlink{0000-0003-2199-0958}
        }
\else
\fi


\ifcase\authorStyle
    \author[\TUM]{P.~Wasser~\orcidlink{0009-0004-7650-7307}}
\or
    \author[\TUM]{
        \fnm{P.} 
        \sur{Wasser} 
        \emailAdd{\newline Wasser, P.: philipp.wasser@tum.de}  
        \orcidlink{0009-0004-7650-7307}
        }
\else
\fi

\ifcase\authorStyle
    \author[\TUM]{A.~Wex~\orcidlink{0009-0003-5371-2466}}
\or
\author[\TUM]{
    \fnm{A.}
    \sur{Wex}
    \emailAdd{\newline Wex A.: alexander.wex@tum.de}
    \orcidlink{0009-0003-5371-2466}
}
\else
\fi

\ifcase\authorStyle
    \author[\TUM]{L.~Wienke~\orcidlink{0009-0006-5548-2109}}
\or
    \author[\TUM]{
        \fnm{L.} 
        \sur{Wienke} 
        \emailAdd{\newline Wienke, L.: lars.wienke@tum.de}  
        \orcidlink{0009-0006-5548-2109}
        }
\else
\fi

\ifcase\authorStyle
\or



    
    









    

    %


    \newcommand{\symAlsoAtCoimbra}{\dag}
    
    
    \newcommand{\symNowAtMPIK}{\ddag}

    \newcommand{\symNowAtLNGS}{\#}


    \newcommand{\symNowAtKiutra}{\textdollar}

    
    
    \newcommand{\symNUCLEUScontactEmail}{*}
     
\else
\fi

\maketitle

\flushbottom

\renewcommand{\thefootnote}{\arabic{footnote}}
\setcounter{footnote}{0}

\section{Introduction}
Particle detectors sensitive to energy deposits of a few tens of eV or below enable the investigation of various physical processes. For instance, direct searches for dark matter particles with masses below 1 GeV~\cite{angloher2023results,hsrl-crvf, alkhatib2021light,EDELWEISS:2019vjv} and experiments studying coherent elastic neutrino-nucleus scattering (CE$\nu$NS~\cite{freedman1974coherent, lindner2017coherent, COHERENT:2017ipa}) with neutrinos from reactors~\cite{nucleus_commissioning, Ricochet:2022pzj, CONUS:2024lnu, CONNIE:2021ggh} require energy thresholds on this scale. Phonon-mediated cryogenic calorimeters~\cite{PhysRevD.96.022009, Cruciani:2022mbb, angloher2023results, Ren:2020gaq, PhysRevD.97.022002}, which offer high sensitivity and total target masses up to $\mathcal{O}$(100~g) or more, are well suited to these experimental goals. In order to achieve the best possible energy resolution, both hardware and software techniques are usually employed.

The NUCLEUS experiment~\cite{Angloher_2019, nucleus_commissioning} 
is designed for the observation of CE$\nu$NS in the fully coherent regime, exploiting the two 4.25 GW$_\text{th}$ reactor cores of the Chooz nuclear power plant in France as $\sim$MeV antineutrino sources~\cite{kluck2022nucleus}. The target detector will consist of gram-scale absorber crystals read out by superconducting transition-edge sensors (TES), featuring energy thresholds of at least $\sim$20~eV on nuclear recoils~\cite{PhysRevD.96.022009, Strauss_2017}. After a commissioning phase at the shallow Underground Laboratory (UGL) of the Technical University of Munich (TUM)~\cite{nucleus_commissioning}, the experiment will soon be relocated to the reactor site at Chooz, where a first technical run is scheduled for 2026.

Although the cryogenic calorimeters developed for NUCLEUS demonstrated reproducible baseline energy resolutions below 10 eV~\cite{PhysRevD.96.022009, PhysRevLett.130.211802, Wex_2025, nucleus_commissioning, Abele_2025}, further improving their sensitivity will ease the CE$\nu$NS detection~\cite{abele2026prospectnucleusexperimentchooz}. In particular, given the spectral shape of the neutrino induced recoil rate, a lower energy threshold means the accessibility to a much stronger signal. In this sense, an important result has been achieved with a custom system to reduce pulse tube vibrations~\cite{Wex_2025}. Here, we present two distinct but not mutually exclusive methods to reach the goal. Each method independently improves the detector sensitivity, while their combination enables reaching even lower energy thresholds. First, we introduce an approach which allows us to identify the TES operating point that provides the best signal-to-noise ratio (SNR), through a detailed scan of the sensor's electrical parameters. Then, we focus on an extension of the optimum filter~\cite{gatti1986processing, radeka1967least}, capable of improving the baseline energy resolution by processing the waveforms of different sensors simultaneously. These optimization procedures have been tested using data recorded by a CaWO$_4$-based calorimeter designed for the NUCLEUS experiment, which constitutes one of the detectors to be employed in the technical run at Chooz~\cite{abele2026prospectnucleusexperimentchooz}. After describing the experimental setup in Sec.~\ref{sec:exp_setup}, we will introduce the methods from a theoretical point of view in Sec.~\ref{sec:methods}. In Sec.~\ref{sec:results} we will then present their results and impact on the baseline resolution.

\section{Experimental setup}\label{sec:exp_setup}

\begin{figure}[htbp]
    \centering
    \includegraphics[scale=0.60]{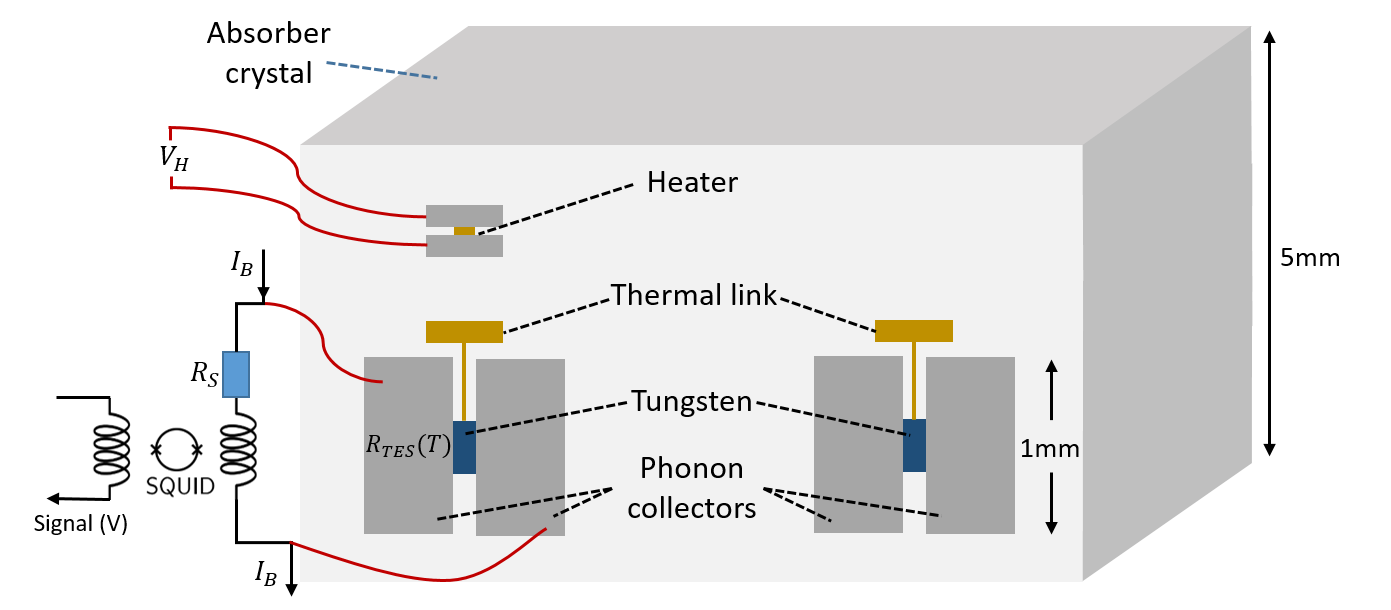}
    \caption{Schematic of the \ce{CaWO4} double-TES detector used in this work. The absorber crystal measures $11.5\times 5\times 5~\mathrm{mm^3}$. The depicted TES size is not to scale. The TES readout circuit is explicitly shown for the left TES but also present for the right one. The TES acts as a temperature dependent resistor $R_{TES}$ and is operated in parallel with a SQUID and a shunt resistor $R_S = \mathrm{40}~ \mathrm{m}\Omega$. Resistance changes in the TES lead to a different branching of the bias current $I_B$, which is detected and amplified by the SQUID and converted into a voltage signal. Only one heater is operated during measurements to heat up the absorber crystal, while the other one is left unconnected.}
    \label{fig:doubleTES}
\end{figure}

The data considered in this work were collected in an above-ground laboratory at TUM. The operated detector was composed of a CaWO$_4$ absorber crystal with dimensions $11.5\times 5\times 5~\mathrm{mm^3}$ and mass $1.74~\mathrm{g}$. The crystal was instrumented with two superconducting tungsten TESs with transition temperatures of $14.56 \pm 0.16$ mK and $14.36 \pm 0.14$ mK, respectively. Each TES is thermally connected to the heat bath through a dedicated gold thermal link. Next to each TES, a resistive gold heater was implemented to enable controlled heating of the detector\footnote{In contrast to the electrothermal feedback operation introduced in Ref.~\cite{IrwinTES}.}. This allows the TES to be stabilized at a defined operating point within its superconducting transition. In addition, aluminum phonon collectors are coupled to the sensors to enhance the signal by improving the collection of non-thermal phonons \cite{Irwin1995QETES}. A schematic of the detector setup is shown in Fig.~\ref{fig:doubleTES}. The detector was operated inside a BlueFors LD400 dry dilution refrigerator to reach the cryogenic temperatures required for TES operation.

For energy calibration and monitoring of the detector response, an optical fiber was installed facing the crystal. It delivered 255~nm optical photons from a room-temperature LED to the detector~\cite{DelCastello:2024ehl}. A $^{55}$Fe X-ray source irradiating the absorber crystal served as an additional calibration reference.

A schematic of the circuit used to read out the detector signal is depicted in Fig.~\ref{fig:doubleTES}. The TES is operated in parallel with a superconducting quantum interference device (SQUID), a highly sensitive magnetometer, which detects small current changes and converts them into a measurable voltage signal.

This voltage is read out by the Versatile Data Acquisition System 2 (VDAQ2), which was specifically designed for cryogenic experiments. The system produces a continuous voltage data stream with a sampling frequency set to $50$~kHz.

The readout circuit is biased with a current $I_B$, which is split between the TES and the SQUID branch. Being operated on the edge of the superconducting transition, the electrical resistance of the TES increases after a temperature rise caused by an energy release inside the absorber. This resistance change leads to an increase in the current flowing through the SQUID branch, which is converted into a corresponding voltage signal by the SQUID readout. An energy deposition in the absorber, therefore, causes a characteristic pulse in the data stream, whose shape can be modeled as described in Ref.~\cite{Probst:1995hjq}.

The detector used here features a double-TES design, incorporating two independent TESs on the same absorber. This configuration was originally developed in Refs.~\cite{CRESST_dbltes,TESSERACT_dbltes} to investigate the low energy background by distinguishing events occurring in the absorber from those originating in the sensors themselves. Each TES is equipped with its own resistive heater, deposited on the crystal surface, so that the heater directly warms the crystal itself. To prevent cross-talk, only a single heater is operated to tune and stabilize both TESs in their respective operating points, while the other one is left unused.

\section{Methods}\label{sec:methods}

\subsection{Operating point optimization strategy}\label{sec3}

\subsubsection*{TES responsivity and noise}

To help contextualize the upcoming discussion, we briefly summarize the relevant aspects of TES responsivity and noise. A much more extensive treatment of TES physics is given in Ref.~\cite{IrwinHilton2005TES}.

A TES is operated within its superconducting transition, where its resistance $R$ is highly sensitive to the temperature $T$. This sensitivity is commonly characterized by the logarithmic temperature coefficient
\begin{equation}
\alpha \equiv \frac{T}{R}\frac{\partial R}{\partial T}.
\end{equation}
The responsivity of a TES describes the conversion of a temperature or energy perturbation into a measurable electrical signal. Importantly, this responsivity is not determined by $\alpha$ alone, but is also modified by the Joule self-heating induced by the bias current and by effects related to the TES critical current. In this sense, the transition is described more generally by a resistance $R(T,I)$ that depends on both temperature and current. The responsivity therefore depends substantially on the operating point (OP), i.e. the depth in the transition and the applied bias current at which the TES is operated.

At the same time, several noise sources contribute to the measured signal. These include thermal fluctuation noise, Johnson noise of the TES and shunt resistor, SQUID noise, as well as additional excess noise (seen in NUCLEUS detectors with $1/f$ dependence). The measured noise is therefore a combination of TES-intrinsic and readout noise, shaped by the electrical and thermal response of the TES. Its dependence on the OP is consequently nontrivial. In this work, we do not attempt to identify or interpret the relative contributions of the individual noise components. Such an analysis would require a more detailed characterization of the noise behavior of the NUCLEUS TES detectors. A dedicated study of these noise contributions may be addressed in future work.

The OP dependance of the signal and noise is illustrated in Appendix \hyperref[appA]{A}, where the signal amplitude and noise level are shown simultaneously across the transition. In particular, the figure illustrates that the signal and noise can exhibit similar relative changes over part of the transition, resulting in broad regions with comparable SNR despite substantially different signal amplitudes.

\subsubsection*{Motivation}

Since the superconducting transition is steep and highly sensitive to thermal conditions, even small changes in either the bias current or the heater power can significantly affect the detector's energy resolution and stability. In the context of NUCLEUS, where the observation of CE$\nu$NS relies on the detection of nuclear recoil energies on the 10 eV scale, achieving the lowest possible energy threshold is crucial. Selecting an appropriate combination of bias current and heater power is therefore a central task in preparing a TES for data taking.

Until now, the OP has typically been selected using simple observables, most notably the amplitude of heater-injected  pulses across the transition. For this purpose, heater sweeps are performed, during which the TES transition is scanned by stepwise reducing the applied heater power. At each heater power setting, a short heater pulse of fixed power is injected on top of the DC heater power, and its amplitude is recorded. The procedure is usually repeated for a few different bias currents. These sweeps provide a rapid overview of the detector’s signal response across its transition. The OP used for data taking is then chosen manually near the maximum of the resulting heater pulse height curve. This fast and pragmatic approach has proven effective in the past, for example in R\&D studies, in determining reliable OPs. 

For the NUCLEUS physics run, however, a more systematic and automatable approach that directly targets optimal detector performance is desirable. In this context, a novel method is presented in which the OP is selected by maximizing the SNR rather than the absolute pulse height alone, thereby explicitly accounting for the detector noise in the optimization. Since the SNR is directly related to the achievable energy resolution, maximizing it corresponds to selecting the OP with the highest sensitivity.

\subsubsection*{Novel Method}

\begin{figure}[htbp]
    \centering
    \includegraphics[scale=0.65]{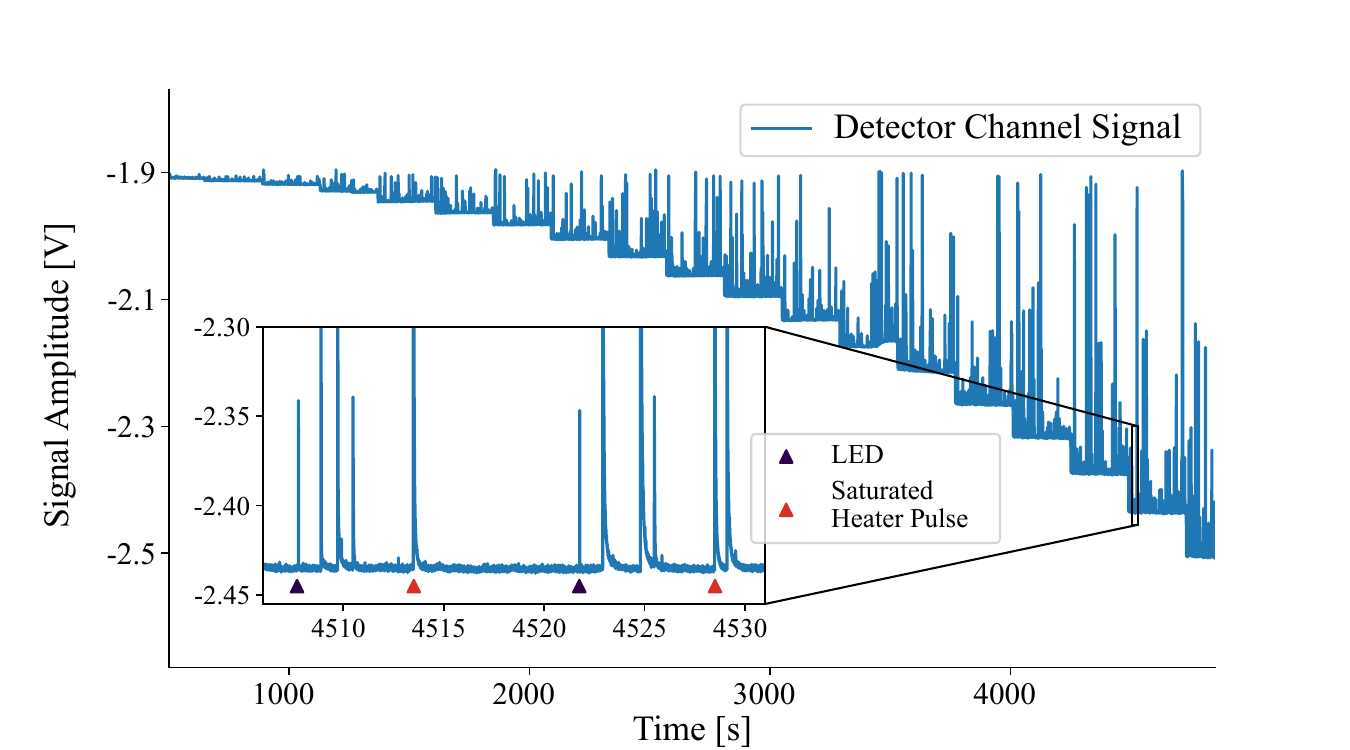}
    \caption{Detector channel response during part of a heater sweep at a fixed bias current of $3.0~\mu\mathrm{A}$. The heater power was decreased every 4 minutes. Each reduction in heater power lowers the detector temperature, resulting in a corresponding decrease of the baseline level in the detector channel. The inset shows artificially injected saturated heater pulses, which are sent every $15$~s to track the depth of the OP on the transition curve. Additionally, LED pulses are sent every $15$~s, which serve as signal reference pulses. Their amplitudes increase as the TES is operated deeper in its superconducting transition. The remaining visible pulses originate primarily from the $^{55}$Fe X-ray source and smaller injected heater pulses.}
    \label{fig:sweep}
\end{figure}

To map the signal and noise conditions over a broad range of OPs, dedicated slow heater sweeps are performed for a set of bias currents between $1.5$ and $4.5~\mathrm{\mu A}$ in steps of $0.5~\mathrm{\mu A}$\footnote{These values correspond to the detector studied in this work. The optimal bias current range and step size depend on the detector design and operating conditions and should therefore be adjusted accordingly.}. For each bias current, the heater voltage is scanned from the normal-conducting regime of the TES through its superconducting transition towards the superconducting state. At each heater setting, the operating conditions are held constant for $4$ minutes, defining a "section" during which the detector can be considered stationary and its performance characterized. An example of such a sweep, recorded for a bias current of $3.0~\mu\mathrm{A}$, is shown in Fig.~\ref{fig:sweep}.

Within each section, two types of reference pulses are injected periodically with fixed and known input settings: (i) saturated heater pulses, which drive the TES out of its superconducting transition and are therefore used to monitor the detector's position within the transition. Their amplitudes are later used to re-establish the selected OP; and (ii) LED pulses, to monitor the detector response: photon bursts of fixed energy are guided to the TES via optical fibers. Energy is deposited in the crystal via electron recoils, generating a thermal pulse. In contrast to heater pulses, the resulting pulse shape closely resembles that of particle interactions, making LED pulses ideal signal references and a key advantage of the presented method.
The LED pulses are injected with identical settings throughout the sweep, ensuring a constant deposited energy and allowing a direct comparison between OPs. Repeating the slow sweep for multiple bias currents yields a two-dimensional grid of OPs in the parameters heater voltage $V_H$ and bias current $I_B$. This dataset forms the basis for the SNR-based OP optimization.

An automated analysis tool, developed specifically for the method, treats each heater–bias combination (i.e. each section) as an independent OP. For every section, a pulse template is constructed by averaging the recorded LED pulses after removing artifacts and pile-up events using a combination of standardized and statistical cuts\footnote{A more detailed description of the applied cuts is provided in Appendix \hyperref[appB]{B}.}. Likewise, randomly sampled noise traces from the corresponding detector stream are cleaned using dedicated data-quality cuts, leaving only pure noise traces from which a noise power spectrum (NPS) is determined. Using the pulse template together with the corresponding NPS, an individual optimum filter (OF)~\cite{gatti1986processing} is constructed for each section. In Fourier space, the OF is given by the ratio of the pulse template to the NPS, weighting each frequency component according to its signal content and noise contribution. This frequency filter preserves the pulse amplitude for single event traces and is thus the standard method for precise amplitude evaluation and baseline resolution determination.

The OF is then applied to the LED pulses and to the noise traces of the corresponding section. We define the median height of the OF-filtered LED pulses as the detector response to the fixed input energy, while the width of the OF-filtered noise distribution provides the baseline resolution (BLR). The SNR of the section is then defined as the ratio of these two quantities:

\begin{equation}
    \text{SNR}_{\text{section}} = \frac{\text{Median OF LED pulse height}_{\text{section}} \text{(V)}}{\text{BLR}_{\text{section}} \text{(V)}}~.
    \label{eq:snr}
\end{equation}

Repeating this procedure for all sections and bias currents yields a two-dimensional map of the detector SNR. An example is shown in Fig.~\ref{fig:snr_curves}. The final OP is selected from this SNR map while taking a few practical constraints into account. For single-TES detectors, a stable point at or near the global SNR maximum can generally be chosen directly. For double-TES detectors sharing a common heater, such as the detector studied here, the two TESs cannot be tuned independently with respect to the heater power. Consequently, the SNR maps are first determined independently for both TES channels and subsequently aligned with respect to the applied heater power. A compromise OP is then selected that simultaneously provides high SNR values for both sensors. 

A previous work~\cite{Wagner_reinforcement} successfully trained a reinforcement-learning model to guide a detector toward its optimum, enabling continuous parameter tuning and potentially rapid convergence. In contrast, the optimization method presented here probes a fixed grid of operating parameters. By relying on longer noise segments and offline analysis, it benefits from significantly higher statistics, resulting in more robust and less noisy SNR estimates while also enabling the application of the optimum filter for a precise evaluation of detector performance. Furthermore, using LED-induced pulses instead of heater-induced pulses provides signal templates with a pulse shape much closer to that of physical particle interactions, making the optimization more representative of detector performance under real operating conditions.

\subsection{2D optimum filter formalism}
\label{sec:2dof}
\begin{figure}[htbp]
    \centering
    \includegraphics[scale=0.6]{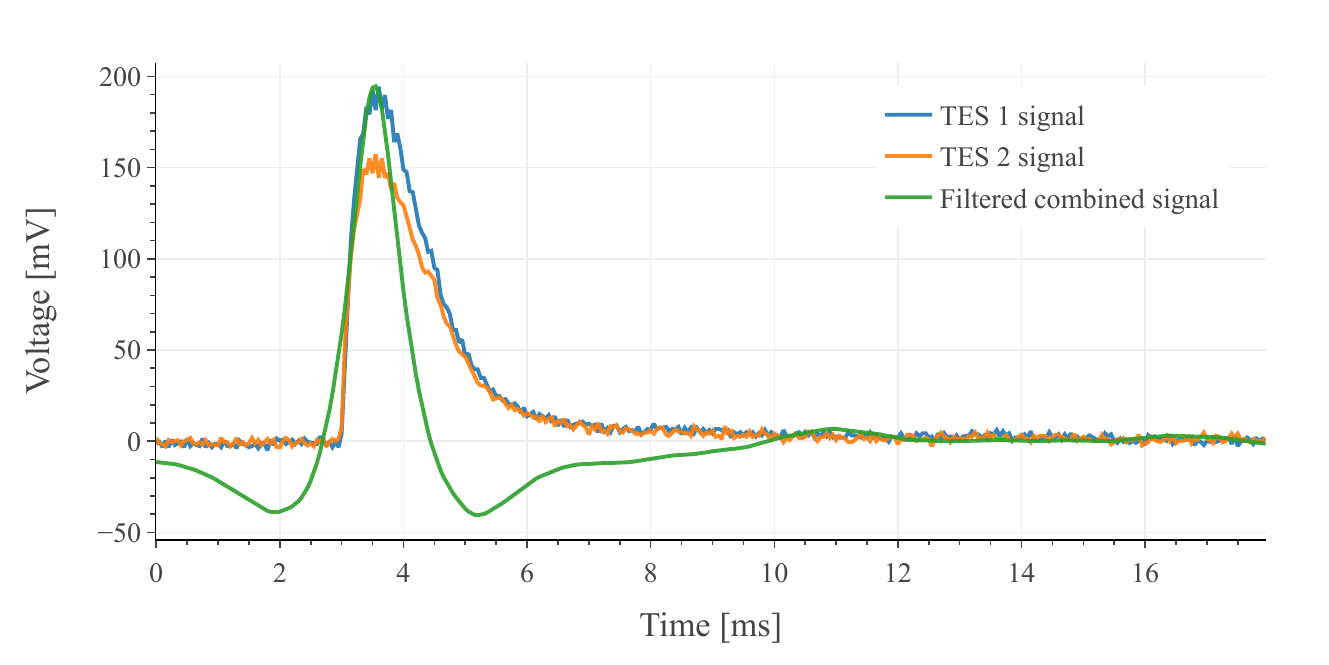}
    \caption{Waveforms on the two sensors (TES~1 and TES~2) following the same energy deposition in the absorber cube, together with the filtered signal in the time domain, obtained from Eq.~\ref{eq:v_filtered}.}
    \label{fig:pulses}
\end{figure}

The double readout of the experimental setup can be exploited to process the data with a two-dimensional optimum filter. The algorithm combines the waveforms in the two TESs, that are simultaneously processed to produce a single filtered signal $v_{\text{filt}}$. In the frequency domain, this can be expressed using a matrix formalism as
\begin{equation}
    \tilde{v}_{\text{filt}}(f) = \frac{1}{K}\, \tilde{S}^\dagger(f) \hat{N}^{-1}(f) \tilde{V}(f)\,,
    \label{eq:v_filtered}
\end{equation}
in which the following quantities have been defined:
\begin{equation}
    \tilde{S}(f) = \begin{vmatrix}
\tilde{s}_1(f) \\
x\tilde{s}_2(f) \\
\end{vmatrix}\,, \quad \tilde{V}(f) = \begin{vmatrix}
\tilde{v}_1(f) \\
\tilde{v}_2(f) \\
\end{vmatrix}\,, \quad \hat{N}_{ij}(f) = \langle \tilde{n}^*_i(f) \, \tilde{n}_j(f)\rangle\,.
\label{eq:definitions}
\end{equation}
In Eq.~\ref{eq:definitions}, $\tilde{S}(f)$ is the column vector containing the Fourier transforms of the template pulses of the two sensors, $\tilde{s}_1(f)$ and $\tilde{s}_2(f)$. Both templates are normalized in the time domain, meaning that $s_1(t)$ and $s_2(t)$ have unit amplitude. The factor $x=A_2/A_1$ is the amplitude ratio of the two template pulses before this normalization is performed\footnote{These normalization conventions ensure that the template pulses are normalized but their amplitude ratio is preserved in the components of $\tilde{S}(f)$. In this way, the filter performs well if the amplitude ratio of the two waveforms is the same of the one of the reference pulses, and any deviation from this normal behavior leads to an incorrect amplitude evaluation.}. In our case the templates are obtained by averaging events that leave signals in coincidence in the two sensors, hence $x$ represents the response ratio to a common energy deposition in the absorber. $\tilde{V}(f)$ contains the waveforms of the two sensors in the frequency domain, and $\hat{N}(f)$ is the $2\times2$ noise covariance matrix, with units of $V^2/Hz$, obtained by averaging over different noise realizations $\tilde{n}_1(f)$ and $\tilde{n}_2(f)$. It is useful to rewrite this last quantity explicitly as:
\begin{equation}
    \hat{N}(f) = \begin{vmatrix}
        N_1(f) & c_{12}(f) \\
        c_{21}(f) & N_2(f) 
    \end{vmatrix}\,,
    \label{eq:noisematrix}
\end{equation}
where $N_1(f)$ and $N_2(f)$ are the noise power spectra of the two TESs while $c_{12}(f)$ and $c_{21}(f)$ are the covariance terms. Finally, $K$ is a normalization constant needed to obtain the correct amplitude when filtering:
\begin{equation}
    K = \int_{-\infty}^{\infty}df\,\tilde{S}^\dagger(f) \hat{N}^{-1}(f) \tilde{S}(f)\,.
    \label{eq:normK}
\end{equation}
The pulse amplitude is directly related to the energy of the event, and is precisely estimated through the filter. The best amplitude evaluation is provided by the maximum of the filtered signal in the time domain $v_{\text{filt}}(t)$. With this formalism, the filtered trace will match the amplitude of the stream of the first sensor.
In Fig.~\ref{fig:pulses}, the waveforms of the two TESs following the same event in the absorber cube are displayed, together with the stream after the filtering procedure. This is an example of the so called "shared events", meaning physical interactions inside the absorber leaving a signal in both sensors. Events of this type are those of interest to us, as opposed to those that leave a signal in only one TES, referred to as "single events".

With respect to the one dimensional optimum filter, this two-dimensional extension provides a lower noise RMS, which translates to the baseline resolution of the detector. A theoretical expectation for this quantity, hereafter referred to $\sigma_{\text{2D}}$, can be computed from the inverse of Eq.~\ref{eq:normK} (following the approach in Ref.~\cite{golwala_thesis}), yielding:
\begin{equation}
\begin{split}
    \frac{1}{\sigma^2_{\text{2D}}} & = \int_{-\infty}^{\infty}df\,\tilde{S}^\dagger(f) \hat{N}^{-1}(f) \tilde{S}(f) = \\ & = \int_{-\infty}^{\infty}df\,\frac{1}{1-|\rho(f)|^2}\Bigg[\frac{|\tilde{s}_1(f)|^2}{N_1(f)} + x^2\frac{|\tilde{s}_2(f)|^2}{N_2(f)} - 2x\operatorname{Re}\Big(\rho(f)\frac{\tilde{s}_1(f)}{\sqrt{N_1(f)}}\frac{\tilde{s}_2^*(f)}{\sqrt{N_2(f)}}\Big) \Bigg]\,,
\end{split}
\label{eq:reso_2d}
\end{equation}
where we introduced the noise correlation between the two sensors
\begin{equation}
    \rho(f) = \frac{c_{12}(f)}{\sqrt{N_1(f) N_2(f)}}\,.
\end{equation}
In the case of completely uncorrelated noise, the resolution of the two-dimensional optimum filter reduces to:
\begin{equation}
    \frac{1}{\sigma^2_{\text{2D}}} = \frac{1}{\sigma^2_1} + \frac{x^2}{\sigma^2_2}\,,
\label{eq:reso_parallel}
\end{equation}
being the standard optimum filter resolution of the $i$-th sensor~\cite{golwala_thesis}:
\begin{equation}
    \frac{1}{\sigma^2_i} = \int_{-\infty}^{\infty}df\,\frac{|\tilde{s}_i(f)|^2}{N_i(f)}\,.
\label{eq:reso_1d}
\end{equation}

\begin{figure}[htbp]
    \centering
    \includegraphics[scale = 0.7]{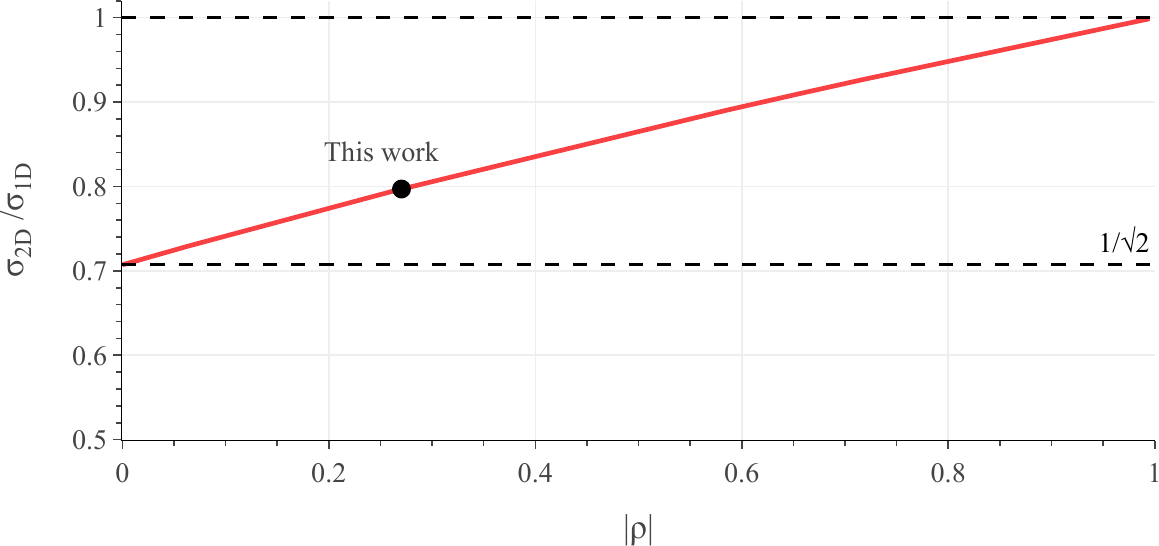}
    \caption{Expected resolution of the two-dimensional OF $\sigma_{\text{2D}}$, computed from Eq.~\ref{eq:reso_2d} in the case of two identical sensors, as a function of the absolute value of the noise correlation $|\rho|$, assuming the phase of the correlation is $\theta=0$. In this simple case, the noise correlation is assumed constant in all the frequency range, and the resolution is normalized with respect to the one-dimensional optimum filter resolution $\sigma_{\text{1D}}$, obtained from Eq.~\ref{eq:reso_1d}. The expected resolution of the NUCLEUS detector employed in this work, given the measured mean values of $|\rho|$ and $\theta$, is marked. The behavior shown here should be considered as a coarse estimate since the non-trivial frequency dependence of the correlation was not taken into account.
    }
    \label{fig:reso_theory}
\end{figure}

The expected resolution from Eq.~\ref{eq:reso_2d} is shown in Fig.~\ref{fig:reso_theory}, in the simple case of two identical sensors with a frequency independent noise correlation defined as $\rho = |\rho|e^{i\theta}$, expressed in terms of its magnitude $|\rho|$ and phase $\theta$. Here the noise is supposed in-phase between the two sensors, i.e. $\theta=0$. The template pulses and noise power spectra have been taken from the actual measurements. A preliminary estimate for the  expected resolution of the detector employed in this work is also marked. This was computed from the mean values (over the frequency range) of the magnitude and phase of the measured noise correlation (presented in Sec.~\ref{sec:2d_application}).

Fig.~\ref{fig:reso_theory} illustrates the mechanism behind the resolution gain of this filtering method. If there is no correlation in the noise phases, the presence of uncorrelated noise results in two independent measurements of the same physical event, enhancing precision. As the noise correlation increases, the two streams become less and less independent, up to the limit in which $|\rho|=1$ and identical waveforms are present on both sensors, with no additional information added by the double readout.

Similar implementations of multi-channels optimum filters for cryogenic particle detectors can be found in Refs.~\cite{hsrl-crvf, ALBAKRY2026171437}.

\section{Results}\label{sec:results}

\subsection{Operating point optimization}\label{sec:results1}

\begin{figure}[htbp]
    \centering
    \includegraphics[scale=0.5]{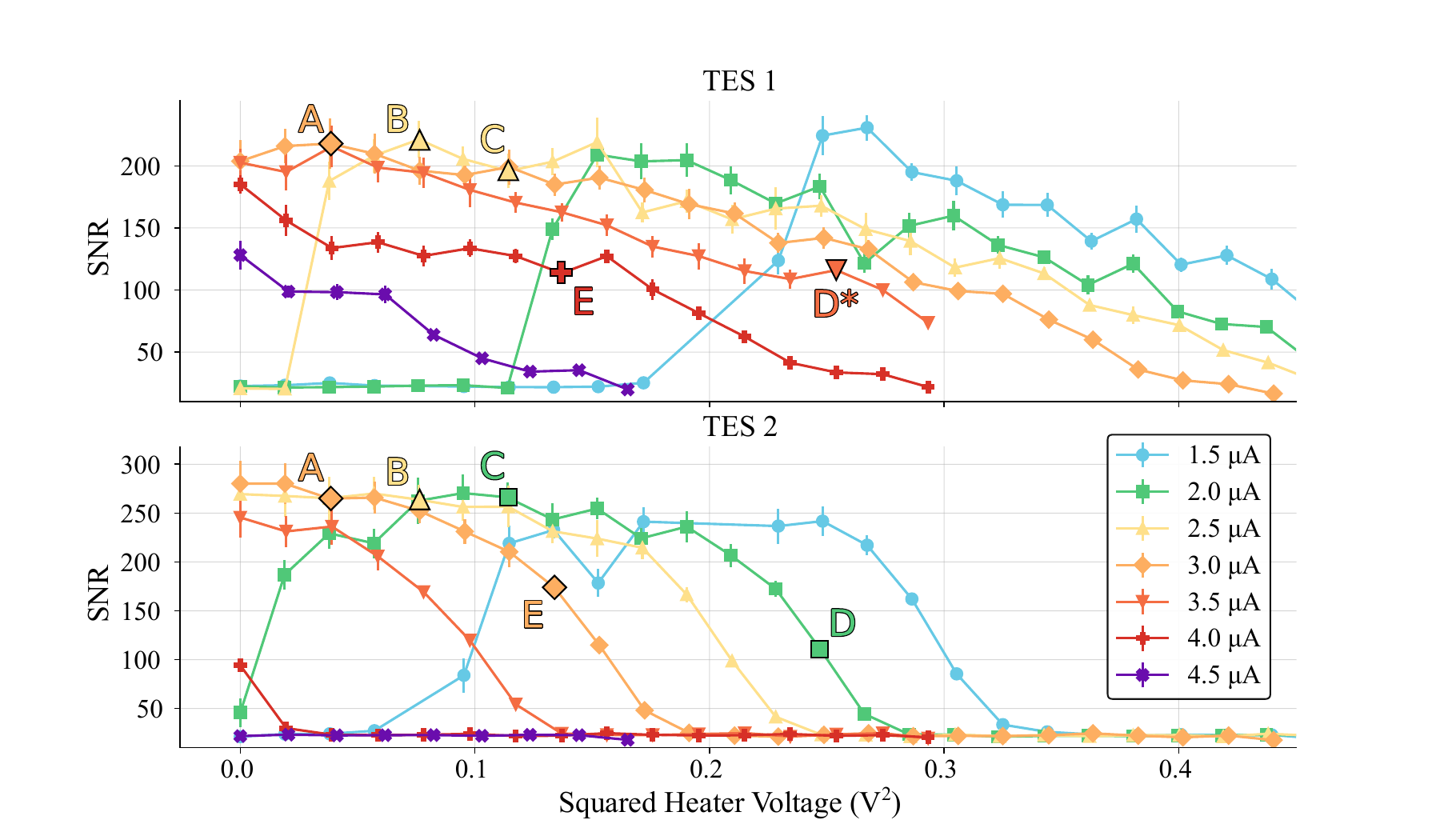}
    \caption{SNR curves obtained from LED pulses for the \ce{CaWO4} double-TES detector. The colors represent the different applied bias currents during the sweeps. Based on this, five different configurations, labeled A-E, were chosen to record calibration data on. For configuration D, the selected OP of TES~1 could not be re-established simultaneously with that of TES~2 during calibration data taking. The position labeled D* indicates the originally expected OP. In practice, the realized OP of TES~1 is shifted to the right toward lower SNR values.}
    \label{fig:snr_curves}
\end{figure}

Slow heater sweeps, described in Sec.~\ref{sec3}, were recorded for the \ce{CaWO4} double-TES detector operated at TUM at bias currents between $1.5$ and $4.5~\mu\mathrm{A}$ in steps of $0.5~\mu\mathrm{A}$, using a section duration of $4~\mathrm{min}$. The corresponding detector stream for a bias current of $3.0~\mu\mathrm{A}$ was shown in Fig.~\ref{fig:sweep}. The automated analysis procedure was then applied independently to both TES channels, yielding the SNR curves shown in Fig.~\ref{fig:snr_curves}. 

The observed behavior can be interpreted in terms of the OP dependence of the TES responsivity and noise, briefly introduced in Sec.~\ref{sec3}. A change in the heater power shifts the position of the TES within its superconducting transition and therefore modifies its responsivity. A change in the bias current affects both the measured signal amplitude and the TES self-heating, thereby shifting its OP. At the same time, the relevant noise contributions also depend on the TES operating conditions. Consequently, increasing the signal response does not necessarily lead to a proportional improvement in the signal-to-noise ratio, since signal and noise change simultaneously as the OP is varied. In particular, over parts of the transition these quantities can change by similar amounts, resulting in a relatively broad range of heater values with maximum SNR, as observed here, rather than a sharply defined optimum. This explains why different combinations of bias current and heater power can yield comparable SNR values despite having different absolute signal amplitudes and noise levels as illustrated in Fig.~\ref{fig:signal_noise} and further discussed in Appendix \hyperref[appA]{A}. \\
Based on the SNR curves, five OP configurations (A--E) were selected. Configurations A--C were chosen from regions where both TESs exhibit high SNR values, while configurations D and E were deliberately placed in less favorable regions to serve as control points. Furthermore, since the SNR exhibits a broad plateau in certain regions, configurations A--C were chosen at different combinations of heater power and bias current with similarly high SNR values. This allows us to test whether comparable SNR values indeed translate into comparable detector performance despite differences in the operating parameters.

Since only one of the two heaters was operated, the TESs cannot be tuned independently with respect to the heater power. Consequently, both TESs share the same heater setting in each configuration. In contrast, no measurable influence of the bias current applied to one TES on the other was observed. The bias currents can therefore be chosen independently for the two sensors, allowing each TES to be operated at its individually preferred bias current while sharing a common heater power. 

To reliably return to the OPs probed during the sweeps, the corresponding saturated heater pulse height was used as a reference for re-establishing each OP. With this approach, all configurations could be reproduced within a few percent of the saturated heater pulse height measured in the corresponding sweep section. The only exception was configuration D, for which the selected OP of TES~1 could not be established simultaneously with that of TES~2, most likely due to a drift in the cryostat temperature during the sweep measurements.

Once the OPs were re-established, extended calibration data were recorded for each configuration. The $^{55}$Fe X-ray source was not considered for calibration, since its energy exceeded the linear range of the detector, leading to saturated pulse shapes. The energy calibration of the two channels was therefore performed using the LED source, sending bursts of optical photons to the crystal through the optical fiber. The detector can be calibrated by exploiting the Poisson statistics of the number of photons reaching the crystal, with a procedure described in Refs.~\cite{ISAILA2012160,Cardani_2018,nucleus_commissioning}. The variance $\sigma^2$ and the mean $\mu$ of the amplitude distribution of events with fixed LED energy are linearly related. Varying the LED injected energy, from a fit of this relation the calibration parameters can be extracted. In this work we do not account for fluctuations in the phonon production, collection or positional effects, which can be a source of systematics of this calibration procedure~\cite{matava}. The systematic uncertainty of the LED calibration, in the case of this detector, is evaluated in another work~\cite{schermer2026developmentcommissioningcryogenictarget} with a direct comparison with the $^{55}$Fe X-ray calibration. This uncertainty, which is expected to be driven by deviations from the detector's linear response and different energy deposition processes between optical photons and X-rays, amounts to $\sim$25\%. A more detailed description of this systematic is provided in Refs.~\cite{nucleus_commissioning,schermer2026developmentcommissioningcryogenictarget}.  

\begin{figure}[htbp]
    \centering
    \includegraphics[width = \linewidth]{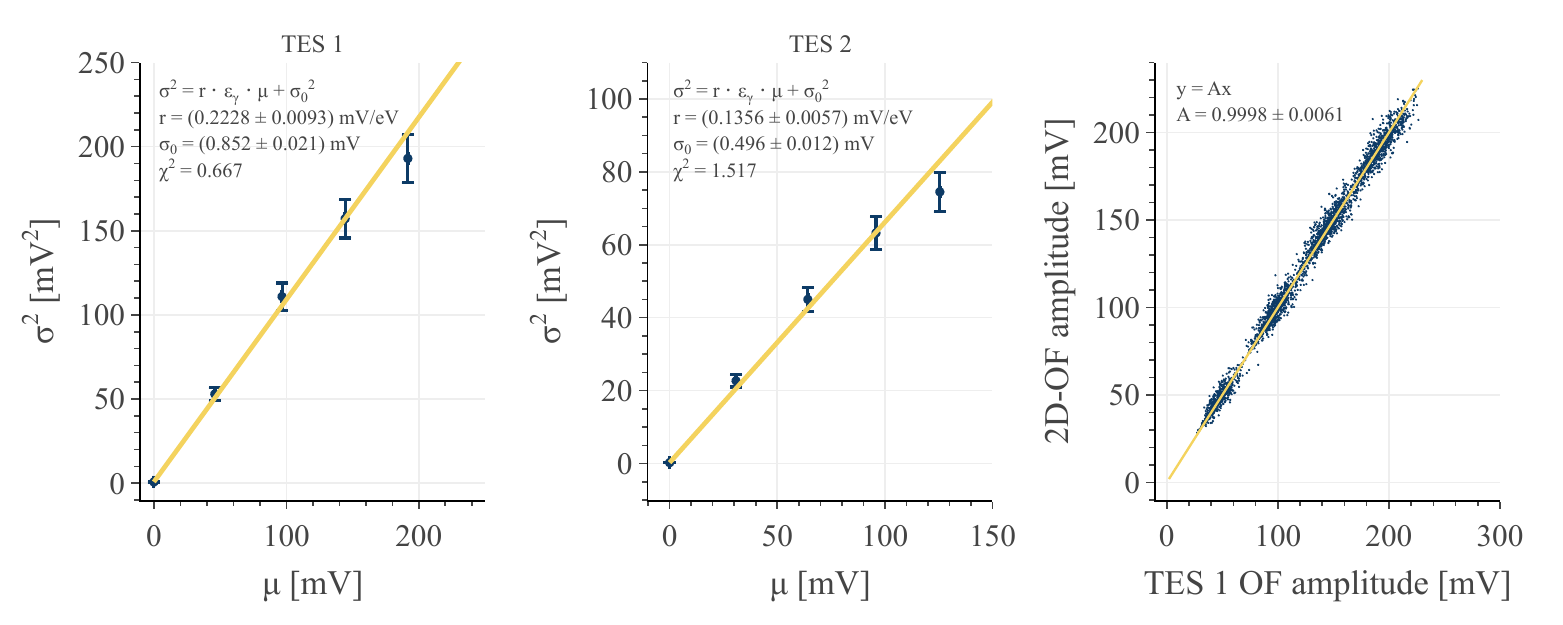}
    \caption{Relation between the variance $\sigma^2$ and the mean $\mu$ of the amplitude distributions evaluated on bursts of photons from the LED source, performed on data from TES~1 (left panel) and TES~2 (middle panel). Fitting with the function $\sigma^2 = r \cdot \varepsilon_{\gamma} \cdot \mu + \sigma_0^2$, where $\varepsilon_{\gamma}$ is the photon energy, the responsivity $r$ and the energy resolution $\sigma_0$ can be obtained as explained in Refs.~\cite{ISAILA2012160,Cardani_2018}. The right panel shows the comparison between two-dimensional and one-dimensional optimum filter (the latter referred to TES~1) in the amplitude evaluation of signals from the LED source. A linear fit was performed to establish the compatibility between the two methods of amplitude estimation. The four different intensities of light bursts from the LED source are clearly visible.}
    \label{fig:calibration}
\end{figure}

\begin{figure}[htbp]
    \centering
    \includegraphics[width=\linewidth]{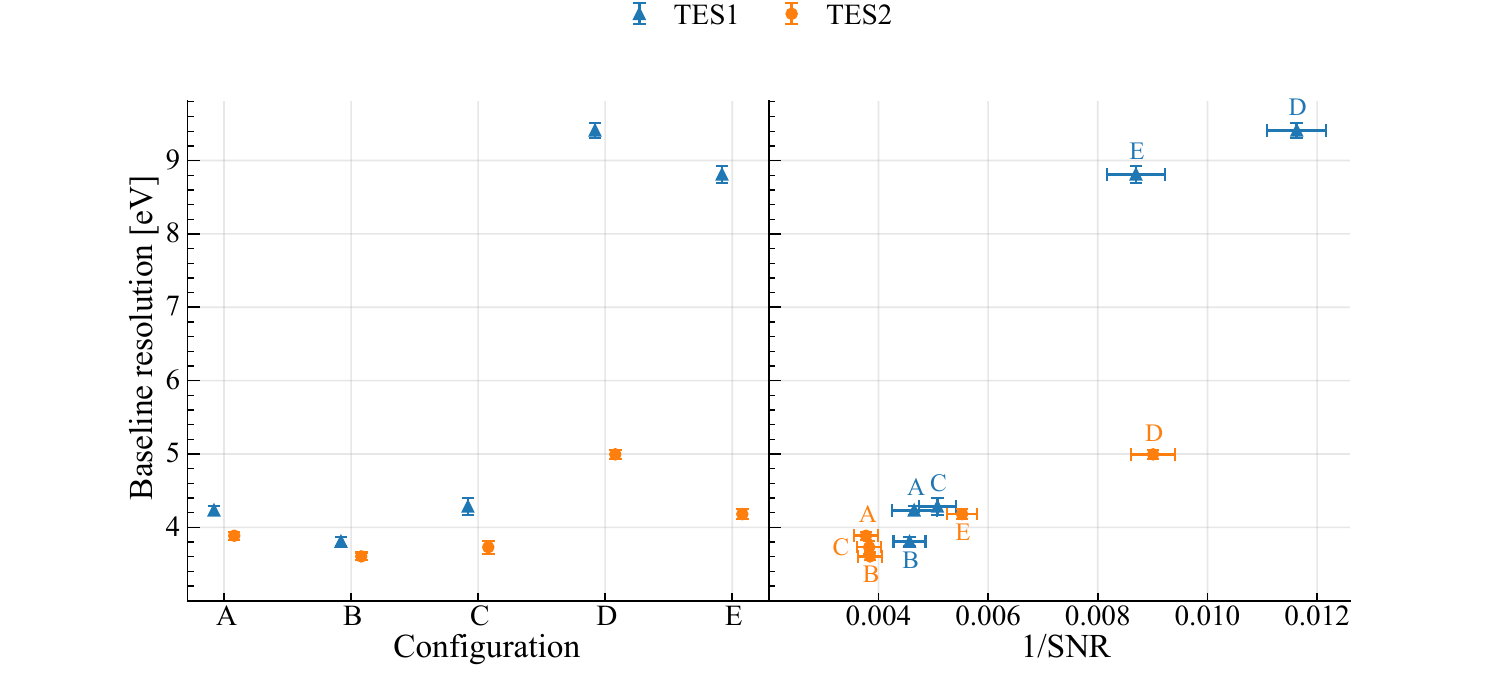}
    \caption{Left: Baseline energy resolutions of the two TESs for the five investigated configurations. The values were obtained by fitting a Gaussian distribution to the amplitudes of filtered noise traces after independently calibrating the two sensors using optical photons. The displayed uncertainties are only statistical. Right: Correlation between the baseline energy resolution measured at the reestablished OPs and the SNR values obtained at the corresponding OPs during the slow heater sweep.}
    \label{fig:resolutions}
\end{figure}

The calibration is performed for every selected configuration, estimating the amplitude of LED pulses with the standard optimum filter for the two channels separately. The energies of LED bursts ranged from $\sim$180 eV to $\sim$1 keV. As an example, the calibration curves for configuration B are displayed in Fig.~\ref{fig:calibration}. The baseline resolution is then obtained performing a Gaussian fit to the amplitude distribution of filtered noise traces, and its values for the different configurations are shown in Fig.~\ref{fig:resolutions}.


These measurements demonstrate that the SNR values extracted from the slow sweeps serve as a reliable indicator of the achievable detector sensitivity. High-SNR configurations result in lower resolution, scaling almost inversely as shown in the right panel of Fig.~\ref{fig:resolutions}. In particular, configuration B achieves the best performance, with both sensors with resolutions below 4 eV. On the other hand, in the low SNR configurations D and E, the resolutions are worse by a factor $\sim$2 for TES~1 and $\sim$1.2 for TES~2.

\subsection{2D optimum filter application}
\label{sec:2d_application}
\begin{figure}[htbp]
    \centering
    \includegraphics[width = \linewidth]{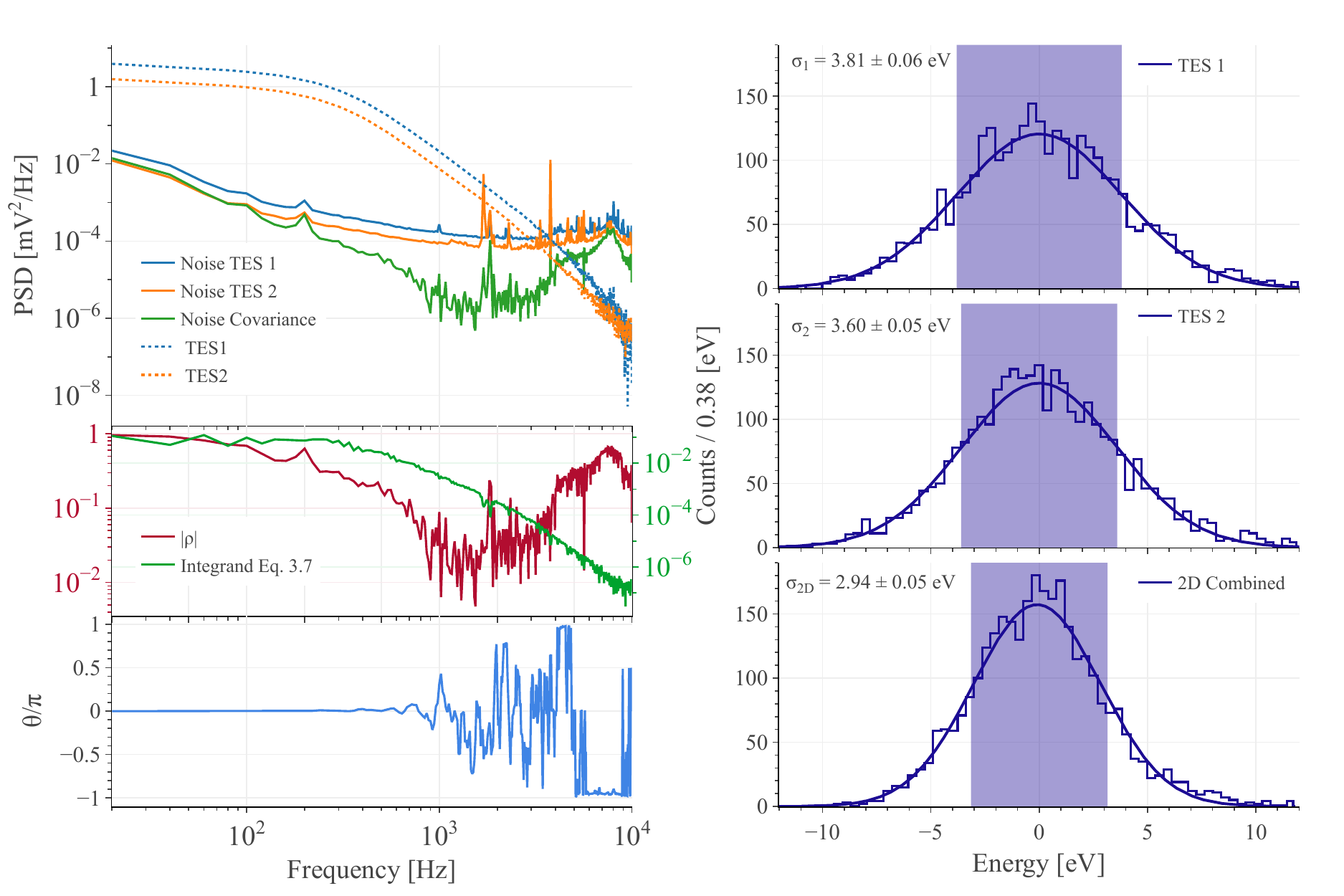}
    \caption{Left: the top panel shows the noise power spectra of the two TESs and the absolute value of the noise covariance, together with the power spectrum of the template pulses. In the middle panel, the absolute value $|\rho|$ of the noise correlation and the integrand of Eq.~\ref{eq:reso_2d} are displayed, while the bottom panel illustrates the phase of the noise correlation $\theta$.  
    Right: histogram of the energy of noise traces obtained with the standard optimum filter on TES~1 (top panel) and TES~2 (middle panel) and with the two-dimensional optimum filter (bottom panel). A Gaussian fit is superimposed to the histograms, and a band of $\pm 1\sigma$ is shown to highlight the width of the distributions. The energy resolutions are also written on the plots.}
    \label{fig:gaussians}
\end{figure}

To study the impact of the procedure described in Sec.~\ref{sec:2dof}, the two-dimensional optimum filter (2D-OF) was applied to calibration data of configuration B. As in the case of the standard optimum filter, the template pulses for the two sensors were obtained by averaging a set of LED pulses. The noise power spectra of the two TESs and the absolute value of the noise covariance $c_{12}(f)$ are displayed in Fig.~\ref{fig:gaussians}, constituting the elements of the noise covariance matrix in Eq.~\ref{eq:noisematrix}. In the same figure, the power spectra of the template pulses, the correlation $\rho(f)$ (in magnitude $|\rho|$ and phase $\theta$), and the integrand of Eq.~\ref{eq:reso_2d} which is the two-dimensional signal-to-noise ratio are also reported. We note that the noise is mostly correlated at low frequencies, where it appears in phase between the two sensors ($\theta=0$), while above $\sim$4 kHz a correlation with noise out of phase in some frequency bands ($\theta \sim \pm\pi$) is visible.

We expect that, when filtering LED events in which the waveforms on the two sensors follow the templates both in shape and in amplitude ratio, the amplitude estimated with the 2D-OF matches the one obtained from the OF. This is indeed what is observed, as displayed in the right panel of Fig.~\ref{fig:calibration}. 

This almost perfect compatibility between the two amplitude estimators allows us to use the same calibration parameter derived from the OF to calibrate the 2D-OF amplitudes. Doing so, we can compare the resolutions and evaluate the gain provided by the 2D-OF (right panel of Fig~\ref{fig:gaussians}). With the two-dimensional filter, we obtain an energy resolution of:
\begin{equation}
    \sigma_{\text{2D}}= (2.94\pm0.05_{\text{stat}}\pm0.74_{\text{syst}}) \, \text{eV}\,.
\end{equation}
This value is smaller than the standard optimum filter resolutions of the two sensors $\sigma_1$ and $\sigma_2$, leading to a gain of $\sim$20\%. We note that this is in line with what we expected from theoretical calculations presented in Fig.~\ref{fig:reso_theory}, where a 20\% reduction was predicted considering two identical sensors with $|\rho| \sim 0.27$ and $\theta \sim 0$. These values represent the mean magnitude and phase of the measured correlation, where the mean is weighted over the entire frequency range. As argued, this procedure takes into account the noise correlation between the sensors. In contrast, by simply averaging the energies on the two TESs, we get an energy resolution of $\sim$3.2 eV.

The smaller baseline resolution provided by the 2D-OF is also advantageous since it should, in principle, translate into a lower trigger threshold. The current trigger scheme in NUCLEUS works by applying a threshold on the data stream, after the latter is filtered with the one-dimensional optimum filter~\cite{Di_Domizio_2011, nucleus_commissioning}. While the effect of a trigger based on the 2D-OF has not been studied yet, we expect a higher SNR at a fixed amplitude used for triggering. Furthermore, the rate of events caused by noise fluctuations (usually called "negative triggers"), which dominate the energy spectrum close to threshold in cryogenic experiments, will be reduced due to the lower noise RMS~\cite{abele2026characterizationlowenergyexcess}.  

\section{Conclusions}
This work presented two distinct techniques for improving the sensitivity of cryogenic calorimeters designed for the NUCLEUS experiment. First, we introduced a method to identify the most sensitive operating points by analyzing short data stream segments and extracting the SNR as a function of heater power and bias current. We demonstrated that operating points with high SNR obtained in this way reliably correspond to the best achievable baseline resolutions. Then, we selected the configuration with best resolution and processed the events with a two-dimensional optimum filter. In this way we were able to further increase the sensitivity, reaching a resolution below 3 eV. 

Beyond the dataset analyzed in this work, we are interested in the impact of these optimizations on an earlier commissioning dataset recorded with an \ce{Al2O3} double-TES detector~\cite{nucleus_commissioning}. Slow sweeps were recorded at the end of the run, identifying an operating point with SNR higher than the one used in the data-taking by (19 $\pm$ 15)\% for TES~1 and (76 $\pm$ 22)\% for TES~2. Although we do not have a direct way to compare the resolutions, since no calibration data were available in this operating point, we can still make a comparison based on some assumptions. The gain provided by the 2D-OF was found to be around 15\%, and we assume here that is independent of the one of the SNR optimization. If the resolution scales linearly with the inverse of the SNR, which is the behavior typically observed, we conclude that we could have reached (3.7 $\pm$ 0.5) eV for TES~1 and (2.7 $\pm$ 0.4) eV for TES~2. 

Taken together, these results strongly show the usefulness of the presented optimization procedure and lay a solid foundation for the technical run at the experimental site in 2026.

\section*{ACKNOWLEDGMENTS}
This work has been financed by the CEA, the INFN, the \"OAW and partially supported by the TU Munich and the MPI f\"ur Physik. NUCLEUS members acknowledge additional funding by the DFG through the SFB1258 and the Excellence Cluster ORIGINS, by the European Commission through the ERC-StG2018-804228 ``NUCLEUS”, by the P2IO LabEx (ANR-10-LABX-0038) in the framework ``Investissements d’Avenir” (ANR-11-IDEX-0003-01) managed by the Agence Nationale de la Recherche (ANR), France, by the Austrian Science Fund (FWF) through the ``P34778-N, ELOISE”, and by Max-Planck-Institut f\"ur Kernphysik (MPIK), Germany. This research was funded in whole or in part by the Austrian Science Fund (FWF) I6955 DOI 10.55776/I6955. 

\newpage

\section*{Appendix A: Illustration of TES Signal and Noise Behavior}\label{appA}

\begin{figure}[htbp]
\centering
\includegraphics[scale=0.50]{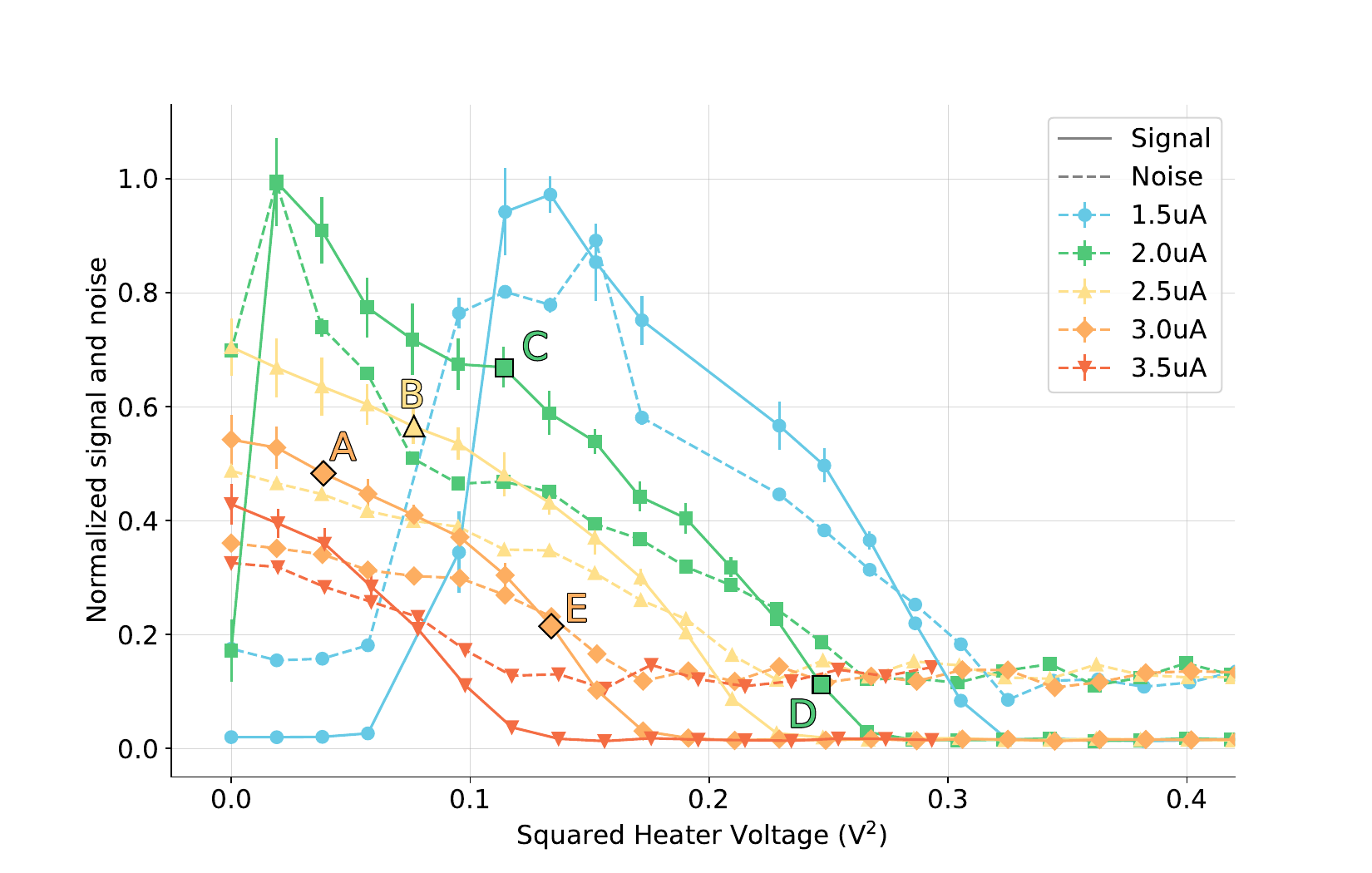}
\caption{Comparison of the signal (solid lines) and noise level (dashed lines) at the investigated OPs for different bias currents for TES~2. Both quantities are normalized to their respective maximum values. The configurations A--E selected for the dedicated calibration measurements are marked on the signal curves.}
\label{fig:signal_noise}
\end{figure}

To further illustrate the OP dependence of the signal and noise underlying the SNR curves displayed in Fig.~\ref{fig:snr_curves}, Fig.~\ref{fig:signal_noise} shows their evolution across the TES transition for different bias currents, using TES~2 as an example. Both quantities are normalized to their respective maximum values to facilitate a comparison of their relative evolution. The signal and noise quantities are determined as described in Sec.~\ref{sec3}.

The signal and noise do not exhibit identical OP dependencies. However, over a substantial part of the transition, their relative evolution is similar, resulting in regions with only weak variation of the SNR and thus explaining the observed SNR plateaus. For example, for a bias current of $2.0~\mu\mathrm{A}$, the signal initially increases more rapidly than the noise as the TES enters its transition, resulting in an increasing SNR. Further into the transition, the signal and noise evolve with similar slopes, leading to an approximately constant SNR. At lower heater powers, the noise increases more rapidly than the signal, causing the SNR to decrease again. This latter behavior is not observed at the higher bias currents shown here, for which self-heating prevents the TES from reaching the same low-temperature regime. The figure also illustrates why the pulse height alone is not a reliable proxy for the SNR. In particular, configurations A--C exhibit similar SNR values despite their different signal amplitudes.

\section*{Appendix B: Automated SNR Analysis Details}\label{appB}

In the following, additional details of the automated analysis procedure introduced in Sec.~\ref{sec3} are provided. The analysis is implemented using the Python-based software package \textit{cait}, developed in the CRESST and COSINUS collaborations and adopted by NUCLEUS~\cite{cait_package}.\\
\textbf{Pulse cuts}: Since the injection times of the LED pulses are known, the corresponding trigger positions are also known. To construct a clean pulse template from the triggers, artifacts such as pile-up events from the $^{55}$Fe source, flux quantum losses (FQLs), and flux quantum resets (FQRs) must be removed. Because some artifacts represent extreme outliers (i.e. an FQR can reach pulse amplitudes of $8~\mathrm{V}$ compared to about $100~\mathrm{mV}$ for a typical LED pulse), we employ the \textit{modified} z-score~\cite{mod_z_score}, which is based on the median and is therefore robust against outliers. We expect the number of such artifacts to be smaller compared to the number of clean LED pulses recorded in each section ($n\approx16$). The modified z-score is applied to both the trace variance and the trace slope\footnote{The trace slope is calculated in cait as the difference between the average at the beginning and the end of the trace.} with a limit of $3.5$ MAD. For the dataset presented here, this procedure yields an average acceptance of approximately $82\%$ in the transition region where LED pulses are observed. \\
\textbf{Noise cuts}: Noise traces are sampled randomly from the continuous data stream and therefore have to be cleaned of any events or baseline distortions. First, a $90^\mathrm{th}$ percentile cut is applied to the RMS of a third-order polynomial fitted to the baseline in order to reject traces with large baseline fluctuations. Subsequently, traces with decaying baselines are removed by applying slope cuts at $\pm3~\mathrm{mV}$. While these thresholds were chosen empirically and may depend on the detector, they have proven suitable for several detectors analyzed with this method. For the present dataset, the combined cuts result in an average acceptance of approximately $86\%$.

\clearpage
\bibliographystyle{JHEP}

\providecommand{\href}[2]{#2}\begingroup\raggedright\endgroup

\end{document}